# Crack arrest markings in stress corrosion cracking of 7xxx aluminium alloys: insights into active hydrogen embrittlement mechanisms

*Martí López Freixes[1], Xuyang Zhou[1], Raquel Aymerich-Armengol[1], Miquel Vega-Paredes[1], Lionel Peguet[2], Timothy Warner[2], Baptiste Gault[1,3]\**

[1] Max-Planck-Institut für Eisenforschung GmbH, Max-Planck-Str. 1, 40237 Düsseldorf, Germany

[2] C-TEC, Constellium Technology Center, Parc Economique Centr'alp, CS 10027, Voreppe, 38341 cedex, France

[3] Department of Materials, Royal School of Mines, Imperial College London, Prince Consort Road, London SW7 2BP, UK

*\* corresponding authors. E-mail addresses: b.gault@mpie.de*

**Abstract**

Crack growth in stress corrosion cracking (SCC) in 7xxx Al alloys is an intermittent process, which generates successive crack arrest markings (CAMs) visible on the fracture surface. It is conjectured that H is generated at the crack tip during crack arrest, which then facilitates crack advancement through hydrogen embrittlement. Here, nanoscale imaging by 4D-scanning-transmission electron microscopy and atom probe tomography show that CAMs are produced by oxidation at the arrested crack tip, matrix precipitates dissolve and solute diffuse towards the growing CAM. Substantial homogenous residual strain remains underneath the fracture surface, indicative of non-localized plastic activity. Our study suggests that H induces crack propagation through decohesion.

Crack-arrest markings (CAMs) are commonly associated with stress-corrosion cracking (SCC) of 7xxx Al-alloys [1,2]. These are generated each time the propagating stress-corrosion crack stops and CAMs illustrates the intermittent crack propagation [3,4]. CAMs are difficult to analyze, as an oxide film often covers the fracture surfaces [2], and there is no consensus regarding processes occurring at the tip during crack-arrest periods and during subsequent crack advancements [5]. The crack growth rates in different alloys and tempers were more dependent on the crack-arrest durations than the distance between CAMs, a critical role of suggesting that processes occurring then — i.e. H production and ingress [2].

CAMs have been proposed to from from oxidation at the crack tip, or by formation of plasticity ridges upon crack arrest [2]. CAM formation is followed by a sudden crack growth, presumably assisted by hydrogen embrittlement (HE) [1,2], through one or more of the proposed mechanisms, including adsorption induced dislocation emission (AIDE) [6], hydrogen enhanced localized plasticity (HELP) [7], or grain boundary decohesion [8]. Other crack growth mechanisms have also been proposed based on mechanical effects, e.g. film-induced cleavage or intermittent increases in local stress-intensity from oxide growth [5].

Here, we aim to address some of the open questions pertaining to the CAMs' chemistry and their structural surrounding. Double cantilever beam (DCB) crack growth tests in hot humid air were performed in specimens extracted from the mid-thickness of a 124 mm 7140 plate. The sample was heat treated to a peak aged state sensitive to SCC. Further information on these aspects can be found in Ref. [9]. We used atom probe tomography (APT), scanning and scanning transmission electron microscopy (SEM & STEM), to further understand the formation mechanism of CAMs, and their role in facilitating crack growth in 7xxx Al-alloys.

We imaged the fracture surface by using a Zeiss Merlin SEM operated at 2 kV and 1 nA. CAMs, Figure 1a, were observed in limited regions [2]. We could not quantify the relative amount CAMs on the fracture surface, as it is mostly covered with oxide following exposure to 85 % RH [9] and CAMs are challenging to image far from the surface (Supplementary Figure 1). The average CAM spacing was 256±127 nm, not accounting for the inclination of the grain boundary plane with respect to the incoming beam. The stress intensity in this region was 22 – 24 MPa√m, and the crack growth rate was ~3 x$10^{-8}$ m/s [9], keeping in mind that the analysis was performed post-mortem. The average crack arrest time, calculated by dividing the average CAM spacing by the crack growth rate, was 9s.

Figure 1b shows a close-up view of the CAMs. The CAM has an oxide-like appearance and in between are of η-phase GB precipitates, indicative of the limited oxidation that has occurred. For simplicity, we use η-phase for η and η', as practically no GP zones are left in at peak age [10]. That the fracture likely occurred close to or directly along the GB plane, as expected for these alloys [1,11]. Voids visible in Figure 1b seem related to η-phase GB precipitates that ended up on the opposite fracture surface [2]. The fractographic evidence does not show clear signs of plasticity, e.g. voids or tear ridges, hinting at a brittle fracture mechanism. However, as argued by Lynch [12], shallow voids may not be resolved by SEM and crack growth involving plastic activity cannot be entirely ruled out.

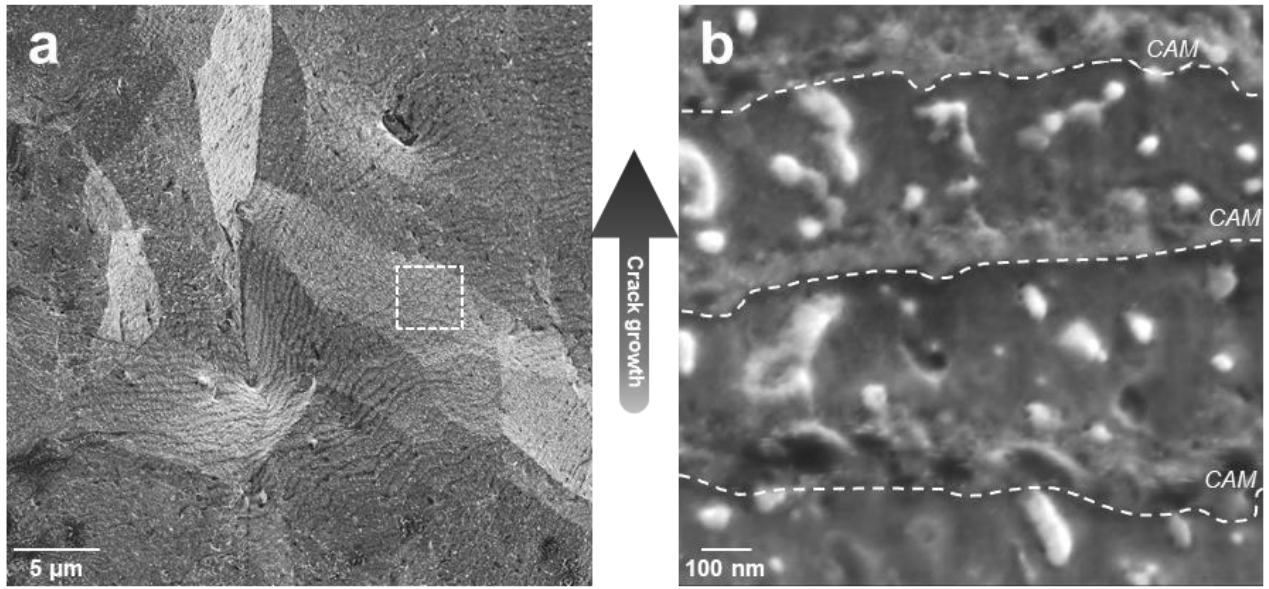

*Figure 1. SEM micrographs showing CAMs on the fracture surface. a) overview of the limited regions where the CAMs were visible. b) close-up view of the CAMs.*

We then turn to STEM and APT analyses (indicative location marked in Supplementary Figure 2). STEM was performed on a ThermoFisher Titan Themis 60-300 equipped with a Cs corrector for the electron probe, operated at 300 kV. Bright field, dark field and high angle annular dark field detectors (collection angles 7 mrad, 10-16 mrad, 18-73 mrad and 78-100 mrad, respectively) were used, along with energy-dispersive X-ray spectroscopy (EDS) with a Bruker Super X-EDX detector. Specimens were prepared following a lift-out method [13] from the fracture surface, using an FEI Helios Xe-PFIB, with a final thinning step at 5kV and 10pA.

The CAMs' cross section in Figure 2a shows the matrix-oxide interface, indicated by a dotted blue line. CAMs appear as semi-elliptical, with an average thickness of 22±13 nm, and are composed of oxidized metal (Supplementary Figure 3). In general, the matrix underneath CAMs does not show signs of intense plastic activity: CAMS are unlikely related to previously suggested plasticity ridges [2]. Small undulations of the interface beneath are observed, marked with red arrows in Figure 2a, although flat, featureless interfaces have also been imaged (Supplementary Figure 4a). Some surface steps are visible, along with void-like features possibly related to the injection of vacancies during oxidation, since they are located below the CAM (Supplementary Figure 4b). Features that could be attributed to plastic activity were observed only under a single CAM, although this might have been caused by the oxide partially falling out of the foil (Supplementary Figure 4c). Between the CAMs, small undulations also appear, with an average height, i.e. the distance between peak and trough, of generally ~5 nm and it not exceeding 8-10 nm (Figure 2a and Supplementary Figure 4a).

APT specimens were prepared using an FEI Helios Xe-Plasma focused ion beam (PFIB) according to the protocol from Ref. [14], following sputter deposition of a Cr-layer to protect the surface (Supplementary Figure 5). APT was performed on a reflectron-fitted Local Electrode Atom Probe (LEAP) 5000 XR (Cameca Instrument Inc), at a base temperature of 80K with a laser pulse energy of 70 pJ at a rate of 200 kHz and with 5 ions detected per 1000 pulses on average. The data was reconstructed and analyzed using AP Suite 6.1 using crystallography-based calibration [15].

Figure 2b displays an APT reconstruction containing part of a CAM, indicated by a dark purple O iso-composition surface, and the interface with the underlying alloy containing η-phase precipitates delineated by a Mg and Zn iso-composition surfaces. Analysis of the nearby matrix η-phase precipitates through the isosurfaces shows that near the CAM, the η-phase precipitates are Zn-rich and Mg depleted, whereas farther away, Mg and Zn are collocated. Comparative composition profiles are plotted in Supplementary Figure 6, illustrating the dissimilar behaviors for Mg and Zn within the precipitates near the CAM.

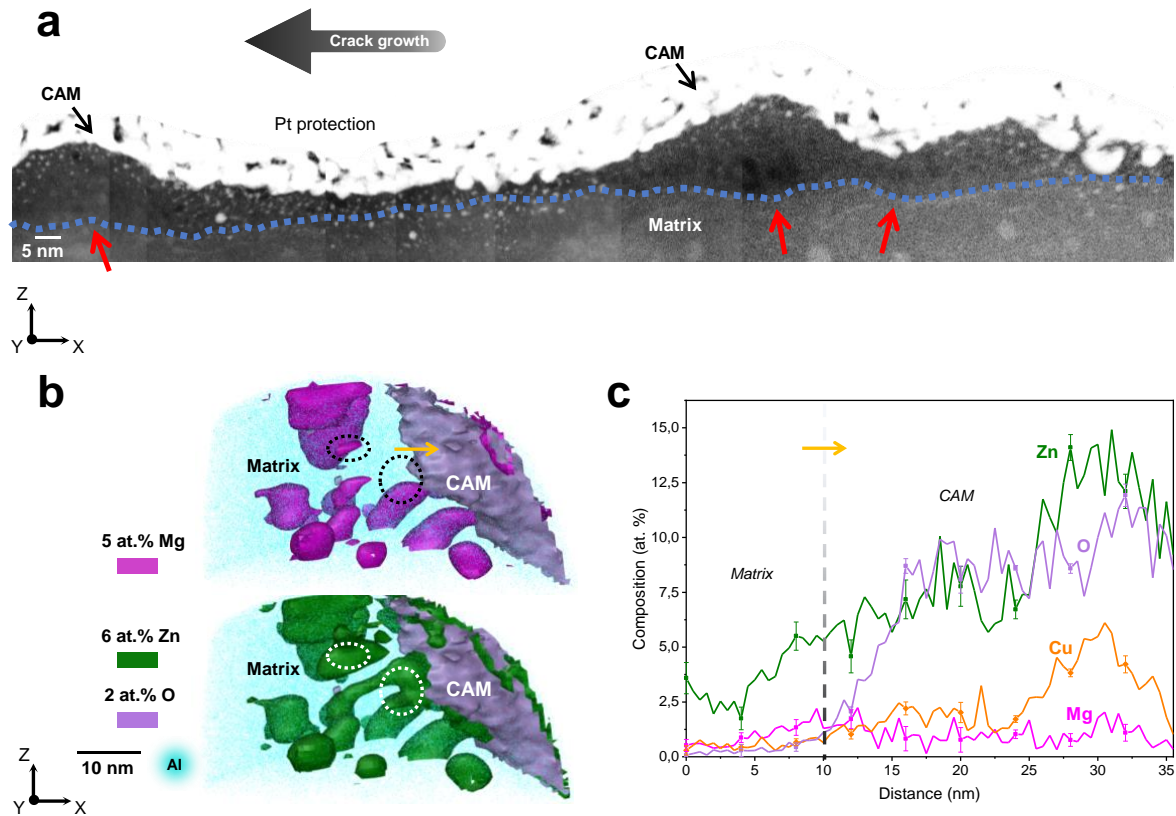

*Figure 2. STEM and APT characterization of the CAMs. a) High-angle annular dark field micrograph showing a cross section view of the CAMs. The interface with the matrix is shown with a dotted blue line, drawn on bright-field micrographs and superimposed here. b) APT reconstruction displaying part of a CAM, shown through an O iso-composition surface. c) Composition profile along the matrix-CAM interface, measured with a 10 nm (ø) cylinder. The error bars correspond to the standard deviation within the bins in the profile. The oxide in the dataset shown in b) was attributed to a CAM based on its size, with a maximum size of ~30 nm along Z. The oxygen-rich feature is not compatible with the oxide layer found on the CAM spacings, which is ~5nm thick as readily visible in a).*

A composition profile along the matrix-CAM interface (Figure 2**Error! Reference source not found.**c) reveals an oxide enriched in Zn, up to approx. 14 at. %, and in Cu, at around 5 at. %. The Mg composition is only approx. 1 at. % on average across the CAM. In contrast, the composition of the oxide at the tip of the crack analysed at lower crack growth rates was higher in Mg, but the Zn content was markedly lower (approx. 3%) [9]. The oxide at the CAM is also hydrated but H levels were removed from the plot for clarity (see Supplementary Figure 7). The adjacent matrix is enriched in Zn, Mg and Cu, with 4.7, 1.4 and 0.7 at. % respectively, averaged over the first 5nm from the CAM as indicated by a dashed line in the composition profile (Figure 2c). This represents a notable increase with respect to matrix compositions far from the crack [9], but also near crack tips that were advancing at lower crack growth rates in the same material, as shown in Table 1. For instance, the Zn content within the matrix adjacent to the CAM is more than double what we reported near the analyzed crack tips (1.93 at. %) [9].

Table 1. Tabulated compositions of the matrix adjacent to a CAM, to the crack tip and far from the crack, i.e. reference.

|  | Zn | Mg | Cu |
|---|---|---|---|
| Matrix adjacent CAM @ 3e-8 m/s | 4.73 ± 0.81 | 1.35 ± 0.33 | 0.66 ± 0.21 |
| Matrix adjacent crack tip @ 3e-10 m/s [9] | 1.93 ± 0.48 | 0.49 ± 0.26 | 0.27 ± 0.17 |
| Matrix reference far from the crack [9] | 0.32 ± 0.04 | 0.27 ± 0.04 | 0.08 ± 0.01 |

Next, we quantified the strain distribution below the fracture surface by using the four-dimensional scanning transmission electron microscopy (4D-STEM) dataset collected by a TemCam-XF416 CMOS detector on a JEM-2200FS TEM (JEOL) operated at 200 kV [16]. The thin foil specimen was the same that was used for STEM imaging. During data acquisition, we precessed the incident electron beam by 0.5° to create quasi-kinematic diffraction patterns [17], and we scanned the sample with a step size of 2.5 nm and a camera length of 25 cm. The data was processed by an open-source code [18] and further analyzed by the py4dstem software package for strain mapping [19].

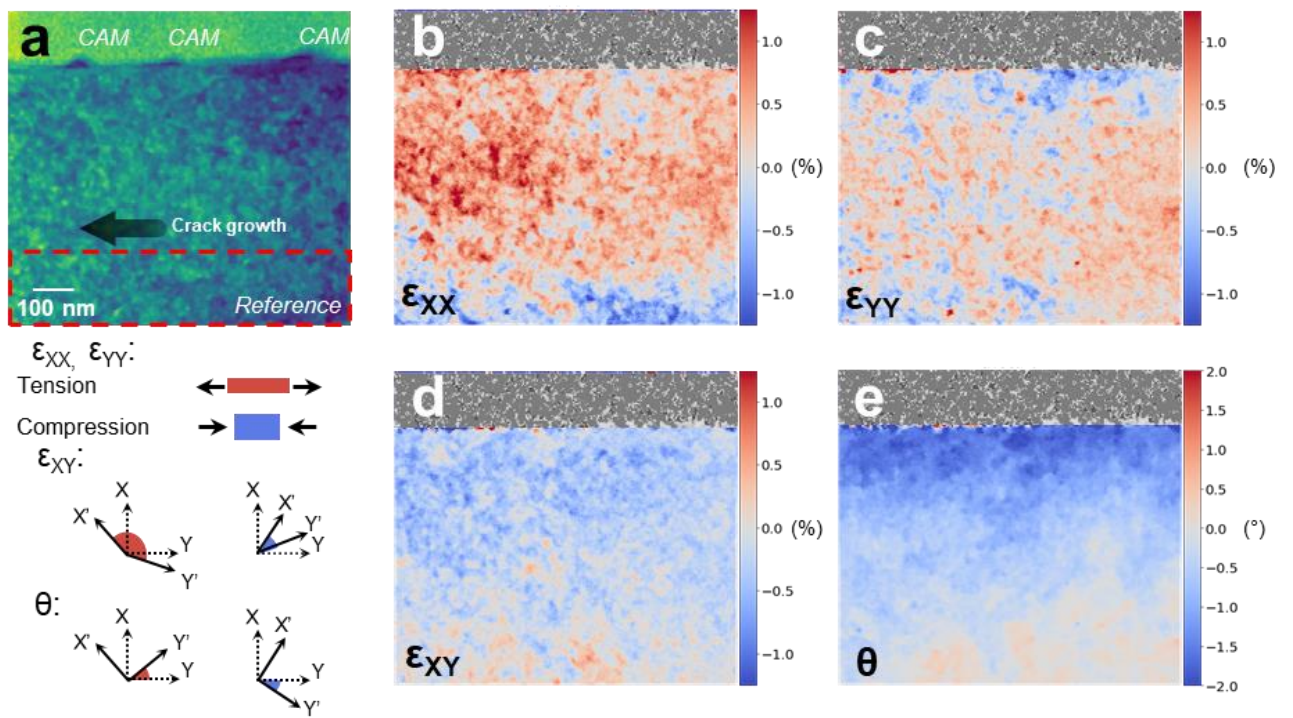

Figure 3. Residual strain measurements below the fracture surface through 4D-STEM. a) Dark field virtual image showin 3 CAMs and the microstructure below them. The reference region for the strain and lattice rotation measurements is shown with a red rectangle and the crack growth direction with a black arrow. Normal strain components are shown in b) $\varepsilon_{XX}$ and c) $\varepsilon_{YY}$. Shear strain component is shown in d) and lattice rotation in e). Grey areas in the images correspond to the Pt protection layer and therefore to unindexed regions.

Figure 3 shows the normal and shear strain maps along with the lattice rotation. The lower region of the scan was used as a reference for the strain calculation (Figure 3a), thus the strain and rotation values are relative to those of the reference. The analysis reveals a rather homogenously distributed tensile strain of 0.4% on average perpendicular to the fracture surface (Figure 3b). The tensile strain in the crack propagation direction is close to zero on average (Figure 3c) and the shear strain component displays a mean value of -0.3% (Figure 3d). There is a non-negligible amount of lattice rotation localized below the fracture surface, up to 2° with respect to the reference region (Figure 3e), which implies the introduction of a significant amount of geometrically necessary dislocations. Rotation of the lattice can be caused by tensile strain [20] but oxidation of an Al pristine surface has also been reported to introduce severe rotations below the oxide [21]. There is no apparent pattern between the position of the CAMs and the strain or lattice rotation maps below the fracture surface.

To summarize, based on our unique combination of experimental insights, we have revealed chemical and structural details of CAMs, along with the residual stress state below the stress-corrosion fracture surface. The CAMs correspond to oxide ridges approx. 20 nm thick, with atom-probe observations indicating that this oxide is solute rich. The matrix adjacent to the CAMs is enriched in solute and Mg depletion within the matrix η-phase precipitates is also observed. 4D-STEM measurements revealed the presence of residual strain in the underlying matrix, the largest strains being tensile perpendicular to the fracture surface although shear strain and lattice rotation are also measured.

Building on the literature, we propose a mechanism to explain the intermittent SCC crack growth and CAM formation schematized in Figure 4. As the crack is arrested water condenses at the sharp crack tip, oxidation quickly begins and a CAM starts to grow (Supplementary Figure 3). Mg from the matrix and also from the GB ahead readily oxidizes, triggering the oxidation-induced dissolution of neighboring matrix η-phase precipitates [22], as suggested by the precipitate vestiges observed near the CAM (Figure 2b and Supplementary Figure 6). The released solutes would then diffuse towards the growing CAM where, due to differences in oxidation rates, Zn and Cu will tend to accumulate at the CAM-matrix interface (Figure 2c) as both Mg and Al would be oxidized beforehand [23]. However, Zn and Cu are eventually oxidized and incorporated into the oxide (Figure 2c). This, in turn, implies that an Al solid solution rich in Zn and Cu will determine the oxidation behavior at the crack tip beyond the first oxidation stages. Alloying elements in an Al solid solution have been shown to alter its corrosion behaviour in Cl-solutions [24–28] but their possible effect upon oxidation in hot humid air/water is unknown.

It should be clarified at this point that it would be indeed surprising that oxidation-induced dissolution of matrix η-phase precipitates and short-range diffusion to the CAM occurs during the average 9 s during which the crack is arrested. If we consider diffusion over a 10 nm distance within 9 s, the required diffusivity is on the order of $10^{-18}$ $m^2$/s, a 6 order of magnitude increase with respect to diffusion of Zn in Al at 70°C [29,30]. Pipe diffusion reportedly accelerates diffusion by 3 orders of magnitude [31] and vacancy introduction during oxidation and plastic deformation [32] may also be contributing. However, considering the extreme diffusivities required, it is plausible that other processes, such as strain-induced dissolution of matrix η-phase precipitates are releasing solutes ahead of the crack, as recently suggested [9]. Further, HE mechanisms such as AIDE [6] and HELP [7] could be assisting the strain-induced dissolution process. The evidence presented herein therefore suggests that, because of the diffusivities required, some of the processes relevant to SCC in 7xxx Al-alloys may be occurring ahead of the crack, although the exact mechanisms by which precipitates dissolve are unclear.

At the crack tip, as the metal oxidizes to form the CAM, atomic H is produced, which partly diffuses ahead of the crack along the boundary [22] (Figure 4), facilitating a sudden crack advancement. Near-atomic scale analysis of the matrix-oxide interface beneath the CAM spacings revealed the absence of any particular feature associated with the advancement of the crack, suggesting that extensive localized plasticity was not the main crack growth mechanism. If the stress-corrosion crack was mainly growing via dislocation emission from the crack tip, via AIDE, or through void coalescence right ahead, through HELP, more extensive traces of plasticity, i.e. voids, would be expected below the fracture surface [12]. A crack growth mechanism involving the coalescence of nanovoids was proposed [12,33], yet for voids much larger than the undulations we observed (Figure 2). In addition, to the author's best knowledge, the presence of voids consistent with the size of the undulations reported herein on SCC fracture surfaces in 7xxx Al-alloys has not been reported. Since imaging can only be performed post-mortem, there is a possibility that voids or other plasticity-related features were removed due to the oxidation, which initiates on defects on Al surfaces [21]. Our observations strengthen the hypothesis of hydrogen-enhanced decohesion (HEDE) playing a key role in crack propagation [8] and the undulations could then be attributed to the topography of the main GB plane along with precipitates.

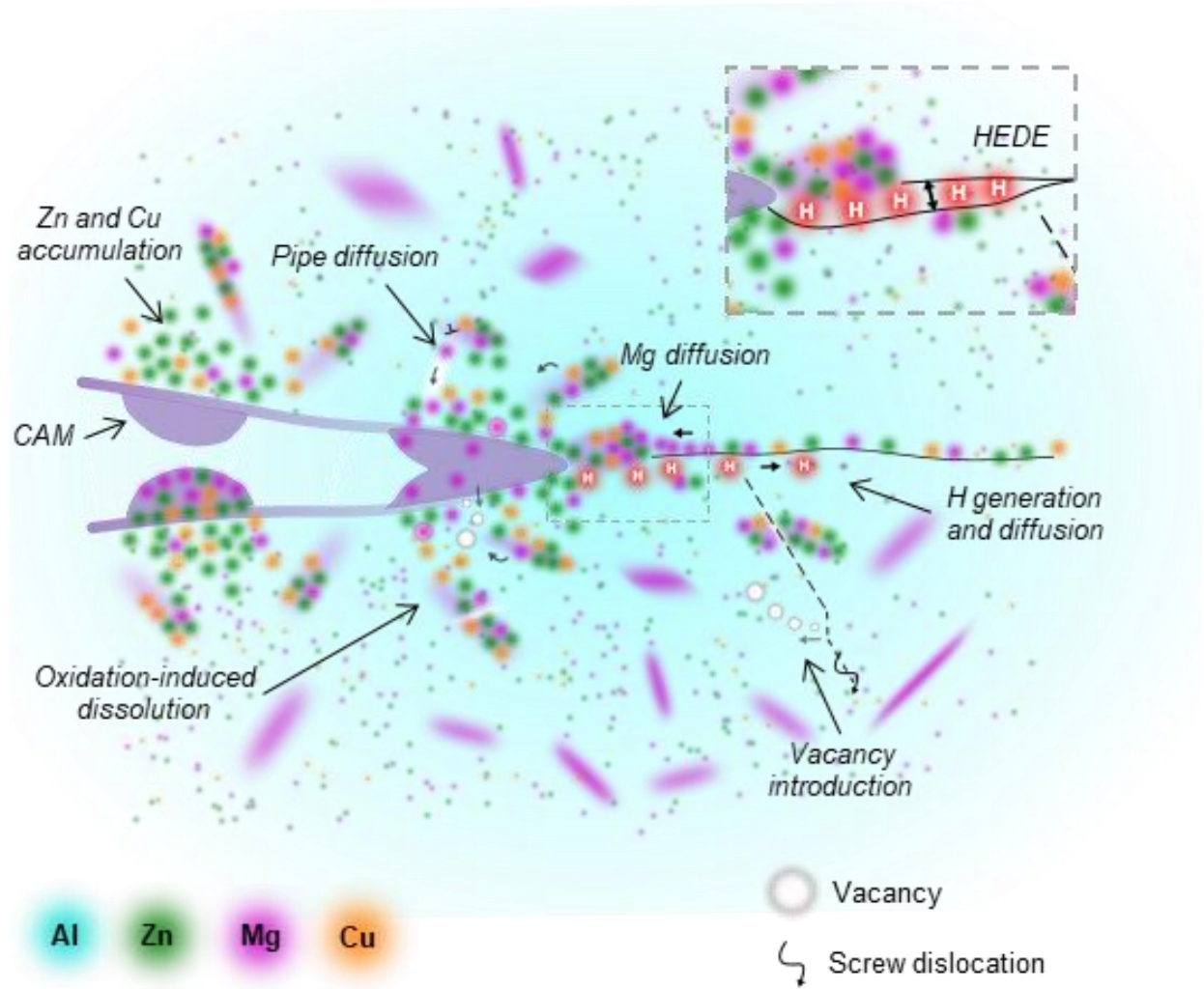

*Figure 4. Interpretation of the evidence and of some the processes that could be at play during SCC of 7xxx Al-alloys.*

Strain mapping beneath the fracture surface through 4D-(S)TEM revealed a substantial amount of residual strain within the foil, which is generated by statistically stored and geometrically necessary dislocations remaining in the foil. Is should be noted that significant relaxation is expected in the thin sample but additional strain could be introduced due to e.g. bending of the foil. The 0.4% average residual $\varepsilon_{xx}$ strain in the foil amounts to ~280 MPa relative to the reference region, calculated through a Hookian law. This could be an indication that significant but seemingly homogeneous plastic activity is occurring during SCC of 7xxx Al-alloys, as opposed to dislocation activity being constrained to the first few nm below the crack. This is also suggested by our STEM observations (Figure 2a), where plasticity features could not be discerned on the matrix-oxide interface. As suggested above, mechanisms of HE such as AIDE [6] and HELP [7] could enhance dislocation emission and its slip, thereby assisting with plasticity although seemingly not directly causing crack propagation.

To conclude, we propose that CAMs are formed exclusively through oxidation, with evidence of precipitate dissolution and solute diffusion, and that the main HE mechanism contributing to crack growth in this peak-aged 7xxx alloy involved decohesion of the main grain boundary, i.e. HEDE. The evidence presented above increases the understanding of SCC in 7xxx Al-alloys, although further observations on different alloys and environments are needed to draw more definitive conclusions on the active mechanisms of SCC in 7xxx Al-alloys and help guide future, more environmentally resistant alloys.

# References

[1] N.J.H. Holroyd, G.M. Scamans, Crack propagation during sustained-load cracking of Al-Zn-Mg-Cu aluminum alloys exposed to moist air or distilled water, Metall. Mater. Trans. A Phys. Metall. Mater. Sci. 42 (2011) 3979–3998. https://doi.org/10.1007/S11661-011-0793-X/TABLES/8.

[2] S.P. Knight, Stress corrosion cracking of Al-Zn-Mg-Cu alloys: effects of heat-treatment, environment, and alloy composition, Monash University, 2008. https://doi.org/10.4225/03/587460E30D193.

[3] P. Martin, J.I. Dickson, J.P. Baïlon, Stress corrosion cracking in aluminium alloy 7075-T651 by discrete crack jumps as indicated by fractography and acoustic emission, Mater. Sci. Eng. 69 (1985) L9–L13. https://doi.org/10.1016/0025-5416(85)90402-1.

[4] C.A. Loto, R.A. Cottis, Electrochemical Noise Generation During Stress Corrosion Cracking of the High-Strength Aluminum AA 7075-T6 Alloy, Corrosion. 45 (1989) 136–141. https://doi.org/10.5006/1.3577831.

[5] S.P. Lynch, Progression markings, striations, and crack-arrest markings on fracture surfaces, Mater. Sci. Eng. A. 468–470 (2007) 74–80. https://doi.org/10.1016/J.MSEA.2006.09.083.

[6] S.P. Lynch, Environmentally assisted cracking: Overview of evidence for an adsorption-induced localised-slip process, Acta Metall. 36 (1988) 2639–2661. https://doi.org/10.1016/0001-6160(88)90113-7.

[7] H.K. Birnbaum, P. Sofronis, Hydrogen-enhanced localized plasticity—a mechanism for hydrogen-related fracture, Mater. Sci. Eng. A. 176 (1994) 191–202. https://doi.org/10.1016/0921-5093(94)90975-X.

[8] H. Zhao, P. Chakraborty, D. Ponge, T. Hickel, B. Sun, C.-H. Wu, B. Gault, D. Raabe, Hydrogen trapping and embrittlement in high-strength Al-alloys, Nature. (2022). https://doi.org/10.1038/s41586-021-04343-z.

[9] M. López Freixes, L. Peguet, T. Warner, B. Gault, Nanoscale perspective on the stress-corrosion cracking behavior of a peak-aged 7XXX-Al alloy, ArXiv. (2023). https://doi.org/https://doi.org/10.48550/arXiv.2303.04625.

[10] T. Marlaud, A. Deschamps, F. Bley, W. Lefebvre, B. Baroux, Influence of alloy composition and heat treatment on precipitate composition in Al–Zn–Mg–Cu alloys, Acta Mater. 58 (2010) 248–260. https://doi.org/10.1016/J.ACTAMAT.2009.09.003.

[11] N.J.H. Holroyd, G.M. Scamans, Stress corrosion cracking in Al-Zn-Mg-Cu aluminum alloys in saline environments, Metall. Mater. Trans. A Phys. Metall. Mater. Sci. 44 (2013) 1230–1253. https://doi.org/10.1007/s11661-012-1528-3.

[12] S. Lynch, Mechanistic and fractographic aspects of stress corrosion cracking, Corros.

Rev. 30 (2012) 63–104. https://doi.org/10.1515/corrrev-2012-0501.

[13] L.A. Giannuzzi, J.L. Drown, S.R. Brown, R.B. Irwin, F.A. Stevie, Applications of the FIB Lift-Out Technique for TEM Specimen Preparation, Microsc. Res. Tech. 41 (1998) 285–290. https://doi.org/10.1002/(SICI)1097-0029(19980515)41:4.

[14] K. Thompson, D. Lawrence, D.J. Larson, J.D. Olson, T.F. Kelly, B. Gorman, In situ site-specific specimen preparation for atom probe tomography, Ultramicroscopy. 107 (2007) 131–139. https://doi.org/10.1016/j.ultramic.2006.06.008.

[15] B. Gault, M.P. Moody, F. de Geuser, G. Tsafnat, A. La Fontaine, L.T. Stephenson, D. Haley, S.P. Ringer, Advances in the calibration of atom probe tomographic reconstruction, J. Appl. Phys. 105 (2009) 034913. https://doi.org/10.1063/1.3068197.

[16] J. Jeong, N. Cautaerts, G. Dehm, C.H. Liebscher, Automated Crystal Orientation Mapping by Precession Electron Diffraction-Assisted Four-Dimensional Scanning Transmission Electron Microscopy Using a Scintillator-Based CMOS Detector, Microsc. Microanal. 27 (2021) 1102–1112. https://doi.org/10.1017/S1431927621012538.

[17] R. Vincent, P.A. Midgley, Double conical beam-rocking system for measurement of integrated electron diffraction intensities, Ultramicroscopy. 53 (1994) 271–282. https://doi.org/10.1016/0304-3991(94)90039-6.

[18] X. Zhou, X. Zhou 2021, (2021). https://github.com/RhettZhou/tvipsBlo.

[19] B.H. Savitzky, S.E. Zeltmann, L.A. Hughes, H.G. Brown, S. Zhao, P.M. Pelz, T.C. Pekin, E.S. Barnard, J. Donohue, L. Rangel Dacosta, E. Kennedy, Y. Xie, M.T. Janish, M.M. Schneider, P. Herring, C. Gopal, A. Anapolsky, R. Dhall, K.C. Bustillo, P. Ercius, M.C. Scott, J. Ciston, A.M. Minor, C. Ophus, py4DSTEM: A Software Package for Four-Dimensional Scanning Transmission Electron Microscopy Data Analysis, Microsc. Microanal. 27 (2021) 712–743. https://doi.org/10.1017/S1431927621000477.

[20] G. Gottstein, Physical Foundations of Materials Science, Springer Berlin Heidelberg, 2004. https://doi.org/10.1007/978-3-662-09291-0.

[21] L. Nguyen, T. Hashimoto, D.N. Zakharov, E.A. Stach, A.P. Rooney, B. Berkels, G.E. Thompson, S.J. Haigh, T.L. Burnett, Atomic-Scale Insights into the Oxidation of Aluminum, ACS Appl. Mater. Interfaces. 10 (2018) 2230–2235. https://doi.org/10.1021/ACSAMI.7B17224/SUPPL_FILE/AM7B17224_SI_005.AVI.

[22] M. López Freixes, X. Zhou, H. Zhao, H. Godin, L. Peguet, T. Warner, B. Gault, Revisiting stress-corrosion cracking and hydrogen embrittlement in 7xxx-Al alloys at the near-atomic-scale, Nat. Commun. 2022 131. 13 (2022) 1–9. https://doi.org/10.1038/s41467-022-31964-3.

[23] M. Hasegawa, Ellingham Diagram, Treatise Process Metall. 1 (2014) 507–516. https://doi.org/10.1016/B978-0-08-096986-2.00032-1.

[24] S. Lameche-Djeghaba, A. Benchettara, F. Kellou, V. Ji, Electrochemical Behaviour of Pure Aluminium and Al–5%Zn Alloy in 3% NaCl Solution, Arab. J. Sci. Eng. 2013 391. 39 (2013) 113–122. https://doi.org/10.1007/S13369-013-0876-7.

[25] F. Sato, R.C. Newman, Mechanism of Activation of Aluminum by Low Melting Point Elements: Part 1 — Effect of Zinc on Activation of Aluminum in Metastable Pitting, Corrosion. 54 (1998) 955–963. https://doi.org/10.5006/1.3284817.

[26] F. Sato, R.C. Newman, Mechanism of Activation of Aluminum by Low-Melting Point Elements: Part 2 — Effect of Zinc on Activation of Aluminum in Pitting Corrosion, Corrosion. 55 (1999) 3–9. https://doi.org/10.5006/1.3283964.

[27] T. Ramgopal, G.S. Frankel, Role of Alloying Additions on the Dissolution Kinetics of Aluminum Binary Alloys Using Artificial Crevice Electrodes, Corrosion. 57 (2001) 702–711. https://doi.org/10.5006/1.3290398.

[28] J. Li, J. Dang, A Summary of Corrosion Properties of Al-Rich Solid Solution and Secondary Phase Particles in Al Alloys, Met. 2017, Vol. 7, Page 84. 7 (2017) 84. https://doi.org/10.3390/MET7030084.

[29] A.W. Nicholls, I.P. Jones, Determination of low temperature volume diffusion coefficients in an Al-Zn alloy, J. Phys. Chem. Solids. 44 (1983) 671–676. https://doi.org/10.1016/0022-3697(83)90115-4.

[30] Y.W. Cui, K. Oikawa, R. Kainuma, K. Ishida, Study of diffusion mobility of Al−Zn solid solution, J. Phase Equilibria Diffus. 2006 274. 27 (2006) 333–342. https://doi.org/10.1007/S11669-006-0005-3.

[31] M. Legros, G. Dehm, E. Arzt, T.J.J. Balk, Observation of Giant Diffusivity Along Dislocation Cores, Science (80-. ). 319 (2008) 1646–1649. https://doi.org/10.1126/science.1151771.

[32] H. Mecking, Y. Estrin, The effect of vacancy generation on plastic deformation, Scr. Metall. 14 (1980) 815–819. https://doi.org/10.1016/0036-9748(80)90295-1.

[33] H.E. Hijnninen, T.C. Lee, L.M. Robertson, H.K. Birnbaum, In Situ Observations on Effects of Hydrogen on Deformation and Fracture of A533B Pressure Vessel Steel, JMEPEG. 2 (1993) 807–818.

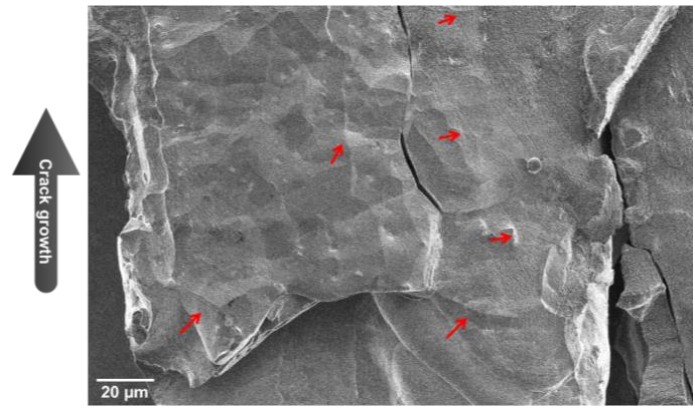

*Supplementary Figure 1. Overview of the intergranular fracture surface. CAMs are marked with red arrows, illustrating the complexity of performing quantitative measurements of coverage %.*

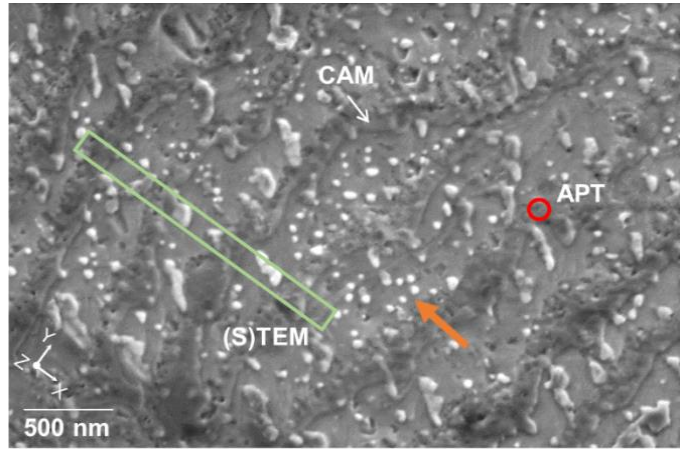

*Supplementary Figure 2. Top view of CAMs on the fracture surface, showing the indicative locations for APT and STEM analysis, along with the crack propagation direction.*

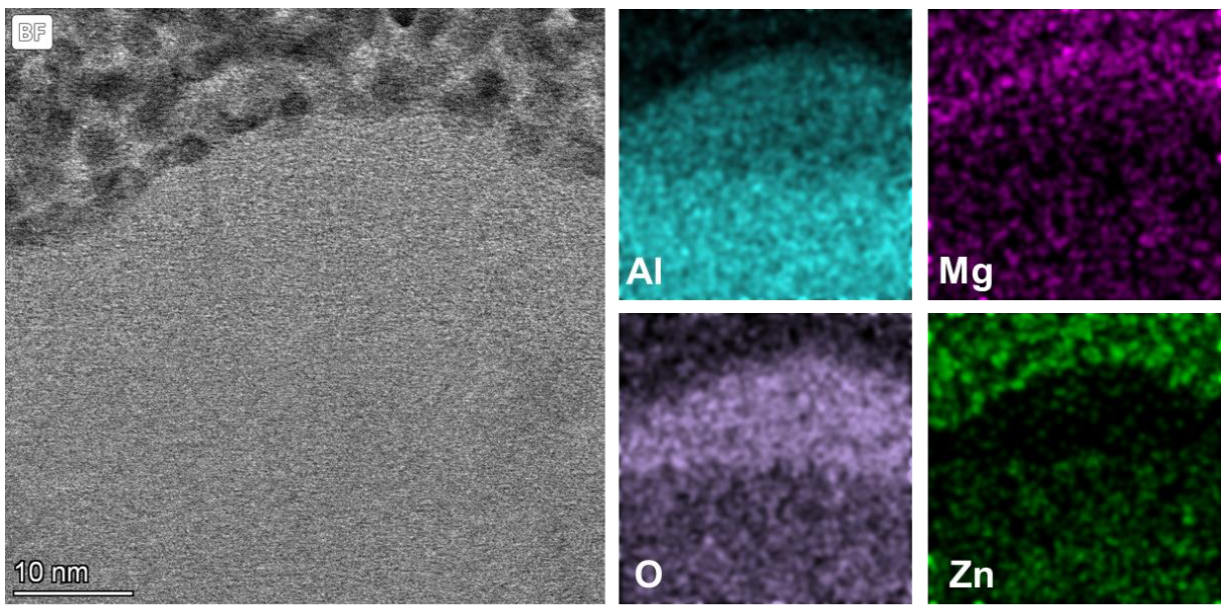

*Supplementary Figure 3. EDS mapping of a CAM.*

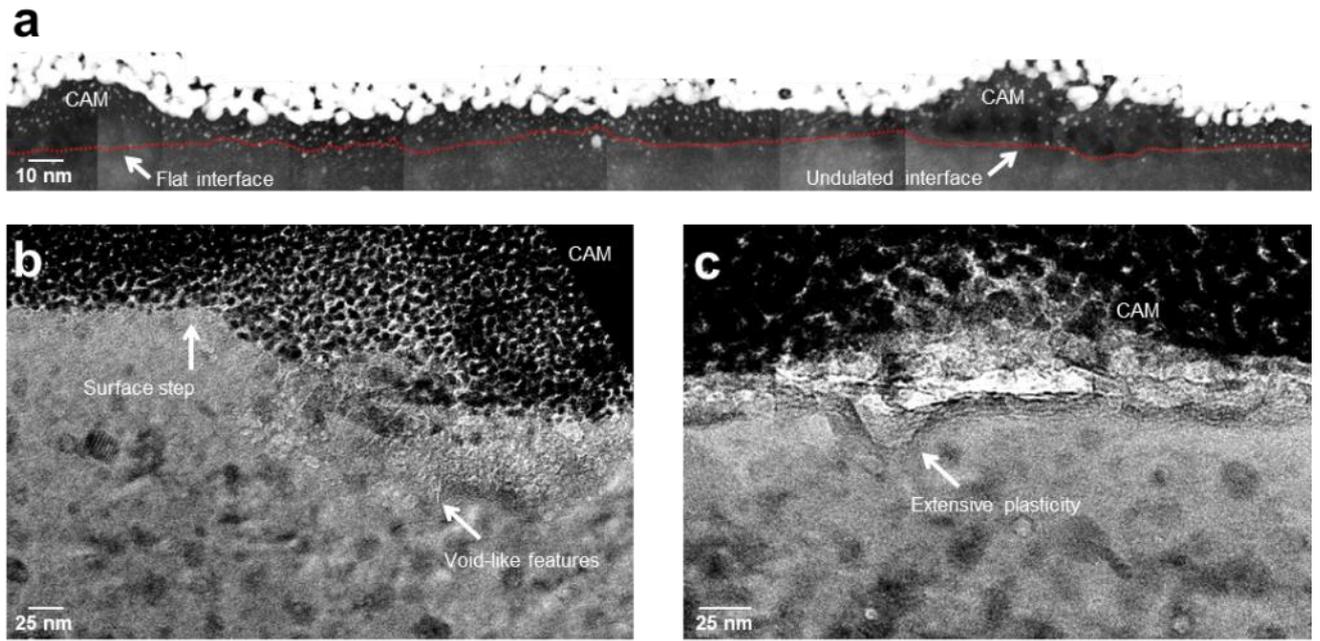

*Supplementary Figure 4. Additional cross-section CAM micrographs showing multiple features.*

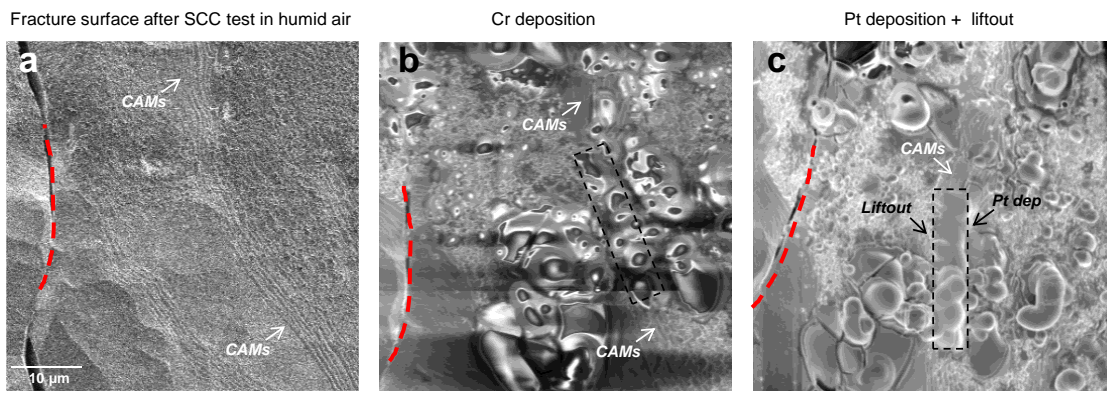

*Supplementary Figure 5. Overview of the sample preparation process for APT CAM analysis*

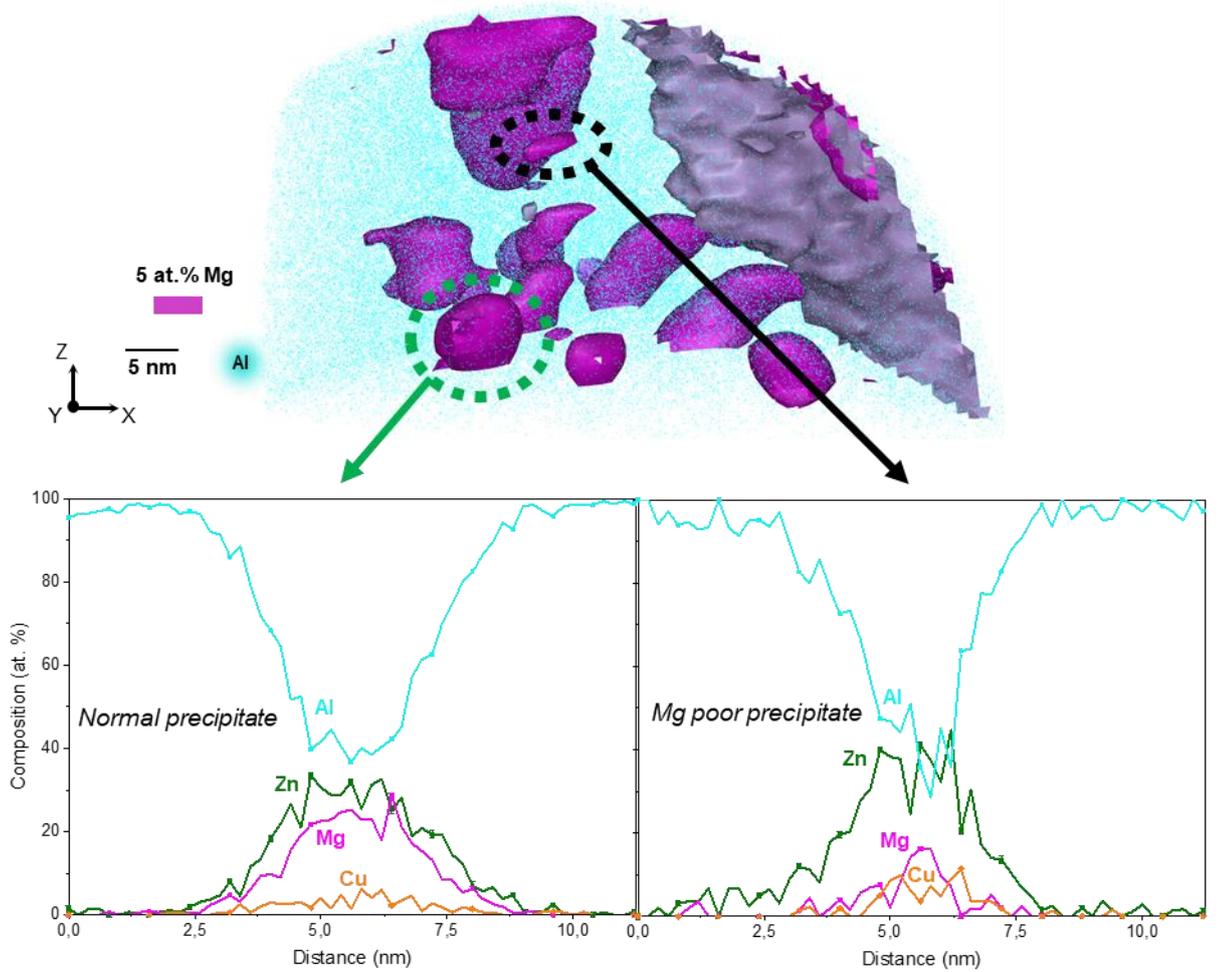

*Supplementary Figure 6. Atom probe dataset containing a CAM (Figure 2) and composition profiles of a precipitate far from the CAM, normal precipitate, and one close to the CAM, poor in Mg.*

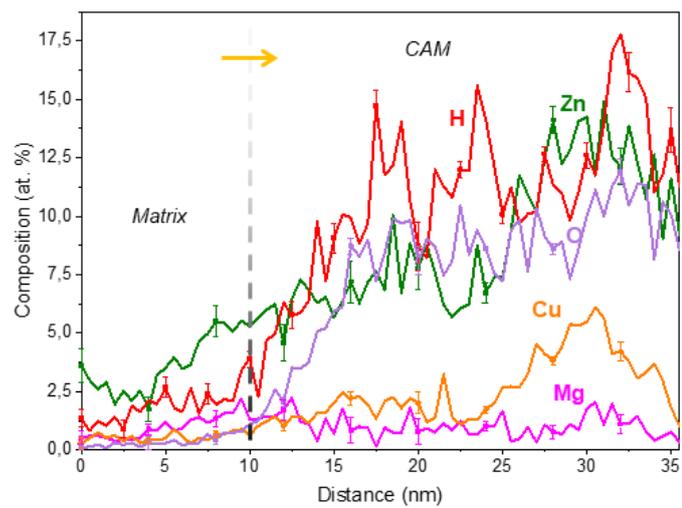

*Supplementary Figure 7. Composition profile along the matrix-CAM interface shown in Figure 2c, with H levels displayed.*

13